\documentstyle[12pt]{article}
\def\be{\begin{equation}}
\def\ee{\end{equation}}
\def\bea{\begin{eqnarray}}
\def\eea{\end{eqnarray}}
\def\real{\hbox{{I}\kern-.2em{\bf R}}}
\begin{document}

\title{CLASS of EINSTEIN-MAXWELL DILATONS, \\
AN ANSATZ FOR NEW FAMILIES OF ROTATING SOLUTIONS }
\author{Tonatiuh Matos$^a$, Dar{\'\i}o N{\'u}{\~n}ez$^b$, Maribel Rios$^c$ \\
%EndAName
$^a$Departamento de F{\'\i} sica, \\
Centro de Investigaci{\'o}n y de Estudios Avanzados del I. P. N.,\\
A. P. 14-700, 07000 M{\'e}xico, D.F.,MEXICO\\
$^b$Instituto de Ciencias Nucleares, \\
Universidad Nacional Aut{\'o}noma de M{\'e}xico\\
A. P. 70-543, 04510 M{\'e}xico, D. F., MEXICO \\
$^c$ Instituto de F{\'\i}sica y Matem{\'a}ticas \\
Universidad Michoacana de San Nicol{\'a}s de Hidalgo \\
Apdo. Postal 2-82, 58040 Morelia, Michoac{\'a}n, M{\'e}xico}
\date{\today}
\maketitle

\begin{abstract}
{The functional potential formalism is used to analyze stationary
axisymmetric spaces in the Einstein-Maxwell-Dilaton theory. Performing a
Legendre transformation, a ``Hamiltonian''is obtained, which allows to
rewrite the dynamical equations in terms of three complex functions only.
Using an ansatz resembling the one used by the harmonic maps ansatz, we
express these three functions in terms of the harmonic parameters, studying
the cases where these parameters are real, and when they are complex. For
each case, the set of equations in terms of these harmonic parameters is
derived, and several classes of solutions to the Einstein-Maxwell with
arbitrary coupling constant to a dilaton field are presented. Most of the
known solutions of charged and dilatonic black holes are contained as
special cases and can be non-trivially generalized in different ways.}
\end{abstract}

\ \ \vspace{7mm} {\bf PACS No. 04.20.--q, 04.20.Fy}

%{\it file armonic01.tex -- 17 de junio de 1998}

\section{Introduction}

Scalar fields as a fundamental interaction in physics are one of the main
predictions of the Kaluza-Klein and the Superstring theories. Scalar fields
are also a fundamental component of the Brans-Dicke theory and of the
inflationary models. Furthermore, in the standard model of Weinberg, Glashow
and Salam scalar fields are needed as a primordial component for given mass
to the particles. More recently, using an exact solution of the
scalar-tensor field equations of gravity, it has been able to show that
scalar fields are a very good candidate to be the dark matter in spiral
galaxies \cite{fco,GMV}.However, it is fair to say that there are still some
open questions in this issue, which should be solved in order to give a more
solid evidence of their presence, and thus establish a proof for the
existence of scalar fields in nature. 

The possibility of existance of scalar fields together with the spontaneous
scalarization in compact stars \cite{dam2} implies that astrophysical
objects could contain scalar fields inherent in them. In other words, if
scalar fields exist, the way it was established in \cite{fco,GMV}, a compact
star will prefer to have one, in order to save energy. Even when these
fundamental scalar fields have not been observed, they are one of the main
ingredients of modern physics. Of course the question arise, why being them
so important in physics have we never seen one? The answer to this question
could be because they interact very weakly with matter. It can be shown that
many of the theories containing scalar fields are in concordance with
measurements in weak gravitational fields \cite{dam,brena}. We expect that
scalar fields are important in strong gravitational fields like at the
origin of the universe or in pulsars or black holes. A great effort has been
done in order to detect scalar fields in strong gravitational fields \cite
{esp}. In some sense, inflation at the origin of the universe could be a
proof of such an interaction. Nevertheless, this effort has been done using
perturbative methods \cite{dam} or using static exact solutions \cite
{MNQ,ma,brena}. The problem with these two approaches is that for the first,
perturbative methods are very efficient only in weak fields and for the
second, astrophysical objects are in general non-static, thus the approach
is not realistic. Even for binary pulsars, perturbative methods have shown a
great success because the distance between the pulsars is so that the
gravitational field is too strong to be understood with Newtonian mechanics
but enough weak to be described by pertubative methods in general relativity 
\cite{esp}. Nevertheless the gravitational interaction is too weak for
deciding which theory containing scalar fields could be the right one. It
has been possible to discard a series of theories which did not agree with
measurements or to bound some parameters of some other theories using
perturbative methods or static exact solutions \cite{brena,dam}. But the
most interesting effects of scalar fields are expected to be very near of a
black hole or of a pulsar and are expected to be non-perturbative. If we
want to understand scalar fields in a strong regime one way to follow is to
find rotating exact solutions of the theory containing scalar fields and
comparing them with observations. The problem then is that the field
equations are very complicated to be solved in an exact manner. In a past
work \cite{MNQ} we gave a very powerful method for finding exact static
solutions of the Einstein-Maxwell-Dilaton field equations using harmonic
maps, we found classes of solutions with arbitrary electromagnetic fields
and gravitational arbitrary multipole momentums. In the present work we
want: i) to give the details of the calculations made in \cite{MNQ}; ii) to
complete the schema of that work, and iii) to present a way to derive exact
rotating dilatons with arbitrary coupling between the scalar field and the
electromagnetic one. In order to do so, let us start from the Lagrangian 
\begin{equation}
{\cal L}=\sqrt{-g}\,(R-2\,(\nabla \,\phi )^{2}-e^{-2\,\alpha \,\phi
}\,F^{2}),  \label{lag0}
\end{equation}
where $g$ is the determinant of the metric tensor, $R$ is the scalar
curvature, $\phi $ the dilaton field and $F$ the Maxwell one. The constant $%
\alpha $ is a free parameter which governs the strength of the coupling of
the dilaton to the Maxwell field. When $\alpha =0$, the action reduces to
the Einstein-Maxwell scalar theory. When $\alpha =1$, the action is part of
the low-energy action of string theory. For $\alpha =\sqrt{3}$, the
Lagrangian (\ref{lag0}) leads to the Kaluza-Klein field equations obtained
from the dimensional reduction of the five-dimensional Einstein vacuum
equations. However, we will consider this theory for all values of $\alpha $.

On the other hand, the harmonic maps ansatz has probed to be an excellent
tool for finding exact solutions of systems of non-linear partial
differential equations \cite{ma24}, in particular, this method has been very
useful in solving the chiral equations derived from a non-linear $\sigma $
model \cite{ma29}. Einstein equations in vacuum can be reduced to a
non-linear $\sigma $ model with structural group $SL(2,R)$ in the space-time
and to a structural group $SU(1,1)$ in the potential spaces, $i.e.$, in
terms of the Ernst potentials. The electro-vacuum case can be also reduced
to a non-linear $\sigma $ model with structural group $SU(2,1)$ in terms of
the extended Ernst potentials \cite{neu}, \cite{kra}. The Kaluza-Klein field
equations can be cast into a $SL(3,R)$ non-linear $\sigma $ model in the
space-time as well as in the potential space \cite{ma1}, \cite{ma24}. This
is possible because the corresponding potential space, defined bellow, is a
symmetric Riemannian space only for $\alpha =0$ and $\alpha =\sqrt{3}$, but
this is not the case for the low energy limit in super strings theory, where 
$\alpha =1$. In this work we will extend the techniques of the harmonic maps
ansatz \cite{neu1}, \cite{ma24}, \cite{mis}, even for non-symmetric
Riemannian spaces, maintaining $\alpha $ as an arbitrary constant. In the
present work, the reduction of the field equations to a non-linear $\sigma $
model is not needed. All through this work, we will be using the MTW \cite
{MTW} signs conventions.

We will analyze spacetimes characterized by two Killing vector fields $X$
and $Y$ and introduce coordinates $t$ and $\varphi $ which are chosen such
that $X={\frac{{\partial }}{{\partial \,t}}}$ and $Y={\frac{{\partial }}{{%
\partial \,\varphi }}}$. The corresponding line element can then be
expressed as 
\begin{equation}
ds^{2}=-f\,(dt-\omega d\,\varphi )^{2}+f^{-1}[e^{2k}(d\rho ^{2}+dz^{2})+\rho
^{2}d\varphi ^{2}],  \label{papa}
\end{equation}
where $f,\ \omega $, and $k$ are functions of $\rho $ and $z$ only. The
electromagnetic potential has the form $A_{\mu }=(A_{0},0,0,A_{3})$, and
again $A_{0},A_{3}$, and the dilatonic field, $\phi $ are functions of $\rho 
$ and $z$ only.

The work is composed as follows: In section 2 we introduce the abstract
potential space, obtain a Lagrangian whose variation with respect to the
potentials reproduce the Field equations. Performing a Legendre
transformation we obtain a ``Hamiltonian'' and finally introduce new
functions which greatly simplify the system of equations. In section 3, by
means of an ansatz resembling the one of the harmonic maps, we express that
system of equations in terms of harmonic parameters, analyzing two cases:
where there are two real harmonic parameters and where there is one complex.
In section 4 we present several new classes of solutions to the
Einstein-Maxwell-Dilaton theory. Finally, in section 5 we present our
conclusions and mention some possibly future developments.

\section{Functional space formulation}

\qquad

{\hspace{1cm}} Trying to solve directly the field equations for the
Einstein-Maxwell-Dilaton theory for the line element (\ref{papa}) can be a
very difficult task. Instead, we will apply the functional geodesic
formulation by defining an abstract space whose coordinates are defined by
the metric functions and the fields entering in the system. In order to
introduce an ansatz resembling the harmonic map ansatz into the functional
geodesic formulation (see \cite{kra}, and \cite{DHA} for an explanation) ,
we shortly explain the general idea of the harmonic map ansatz method. The
field equations of the theory can be written as \bigskip 
\begin{equation}
(\rho \Psi _{\,\,,\varsigma }^{A})_{,\bar{\varsigma}}+(\rho \Psi _{\,\,,\bar{%
\varsigma}}^{A})_{,\varsigma }+2\rho \{_{B\;C}^{A}\}\Psi _{\,\,,\varsigma
}^{B}\Psi _{\,\,,\bar{\varsigma}}^{C}=0  \label{diser}
\end{equation}
where $\varsigma =\rho +i\ z$ and ${\bar{\varsigma}}$ is its complex
conjugated. $\Psi ^{A}$ are the potentials of the geodesic formulation and $%
\{_{B\;C}^{A}\}$ are the Christoffel symbols of the Riemannian space $V^{A}$
defining the potential space of the theory. Now we look for invariant
transformations of the equations (\ref{diser}), $i.e.$, transformations of
the form $\Psi ^{A}=$ $\Psi ^{A}(\lambda ^{i})$ that leave the field
equations (\ref{diser}) invariant, where $\lambda ^{i}$ are potentials
fulfilling the same field equations (\ref{diser}). The potentials $\lambda
^{i}$ define the Riemannian space $V_{p}$. In terms of the potentials $%
\lambda ^{i}$, the field equations (\ref{diser}) read 
\begin{equation}
2\rho \lbrack \Psi _{\,\ \,,i\,j}^{A}+\{_{B\;C}^{A}\}\Psi _{\,\,,i}^{B}\Psi
_{\,\,,j}^{C}]\lambda _{,\varsigma }^{i}\lambda _{,\bar{\varsigma}}^{j}+\Psi
_{\,\,,i}^{A}[(\rho \lambda _{\,\,,\varsigma }^{i})_{,\bar{\varsigma}}+(\rho
\lambda _{\,\,,\bar{\varsigma}}^{i})_{,\varsigma }]=0,  \label{dieser2}
\end{equation}
where $,i={\partial /}\partial \lambda ^{i}$. In terms of the Christoffel
symbols of the abstract Riemannian space $V_{p}$, (\ref{dieser2}) reads 
\begin{equation}
\Psi _{\,\ \,,i;\,j}^{A}+\{_{B\;C}^{A}\}\Psi _{\,\,,i}^{B}\Psi
_{\,\,,j}^{C}=0  \label{dieser3}
\end{equation}
where we have used the field equations for the $\lambda ^{i}$'s and the fact
that the $\lambda ^{i}$'s are linear independent.

In what follows we will introduce the functional geodesic formulation for
Lagrangian (\ref{lag0}). The method is fully explained in \cite{DHA}.
Essentially, the formulation takes advantage of the fact that the
introduction of a line element in the Lagrangian for the Einstein-Hilbert
action, Eq.~(\ref{lag0}), and performing the variation, are operations which
commute for some cases, in particular for the axisymmetric stationary one.
Thus, introducing the operator $D=(\partial _{\rho },\partial _{z})$, taking
out a total divergence term and eliminating the terms with $D\,k$ by means
of a Legendre transformation, we obtain that the original Lagrangian, given
by Eq.~(\ref{lag0}), can be rewritten as: 
\begin{equation}
{\cal L}={\frac{\rho }{{2\,f^{2}}}}\,Df^{2}-{\frac{{f^{2}}}{{2\,\rho }}}%
\,D\omega ^{2}+{\frac{{2\,\rho }}{{\alpha ^{2}\,\kappa ^{2}}}}\,D\,\kappa
^{2}+{\frac{{2\,f\,\kappa ^{2}}}{{\rho }}}\,[(\omega \,DA_{0}+DA_{3})^{2}-{%
\frac{{\rho ^{2}}}{{f^{2}}}}\,DA_{0}^{2}],  \label{lag1}
\end{equation}
where $\kappa ^{2}=e^{-2\,\alpha \,\phi }$.

The Euler-Lagrange equations, obtained directly from extremizing the action
for such Lagrangian: $D({\frac{{\partial {\cal L}}}{{\partial D\,Z^{a}}}})-({%
\frac{{\partial {\cal L}}}{{\partial Z^{a}}}})=0$, with $Z^{a}=(f,\omega
,A_{0},A_{3},\kappa )$, are:

\noindent the Klein-Gordon equation: 
\begin{equation}
D^{2}\,\kappa +({\frac{{D\,\rho }}{\rho }}-{\frac{{D\,\kappa }}{\kappa }}%
)\,D\kappa -{\frac{{\alpha ^{2}\,\kappa ^{3}\,f}}{{\rho ^{2}}}}[(\omega
\,DA_{0}+DA_{3})^{2}-{\frac{{\rho ^{2}}}{{f^{2}}}}\,DA_{0}^{2}]=0,
\label{Klein-Gordon}
\end{equation}
the Maxwell equations: 
\begin{eqnarray}
D\left( {\frac{{f\,\kappa ^{2}}}{{\rho }}}\,(\omega \,DA_{0}+DA_{3})\right)
&=&0,  \nonumber \\
D\left( \kappa ^{2}\,[{\frac{{f\,\omega }}{{\rho }}}(\omega \,DA_{0}+DA_{3})-%
{\frac{{\rho }}{{f}}}\,DA_{0}]\right) &=&0,  \label{maxw}
\end{eqnarray}
and the main Einstein's equations: 
\begin{eqnarray}
&D^{2}\,f+({\frac{{D\,\rho }}{\rho }}-{\frac{{D\,f}}{{f}}})\,D\,f+{\frac{%
f^{3}}{\rho ^{2}}}\,D\,\omega ^{2}-{\frac{{2\,\kappa ^{2}\,f^{2}}}{{\rho ^{2}%
}}}\,[(\omega \,DA_{0}+DA_{3})^{2}+{\frac{{\rho ^{2}}}{{f^{2}}}}%
\,DA_{0}^{2}]=0&,  \nonumber \\
&D^{2}\,\omega -({\frac{{D\,\rho }}{\rho }}-{\frac{{2\,D\,f}}{{f}}}%
)\,D\,\omega +{\frac{{4\,\kappa ^{2}}}{{f}}}\,(\omega
\,DA_{0}+DA_{3})\,D\,A_{0}=0&,  \label{Einstein}
\end{eqnarray}
which are equivalent to the usual field equations. Notice that the metric
function $k$ does not appear, reflecting the fact that it is determined by
quadratures in terms of the rest of the functions. The Lagrangian (\ref{lag1}%
) sometimes is viewed as describing the line element in the potential space,
and thus the motion equations can be seen as geodesics in that space.

From the fact that $D\,{\tilde{D}}=0$ for any analytic function, with ${%
\tilde{D}}=(-\partial _{z},\partial _{\rho })$, we conclude from the second
Maxwell equation, Eq.(\ref{maxw}), that there exits a potential $\chi $,
such that 
\begin{equation}
{\tilde{D}}\,\chi ={\frac{{2\,f\,\kappa ^{2}}}{\rho }}\,(\omega
\,DA_{0}+DA_{3}).  \label{chi}
\end{equation}
And with this potential, the second Einstein's equation, Eq. (\ref{Einstein}%
), can be rewritten as $D({\frac{{f^{2}}}{\rho }}\,D\,\omega +\psi \,{\tilde{%
D}}\,\chi )=0$, with $\psi =2\,A_{0}$, so that there exists another
potential, $\epsilon $, defined by 
\begin{equation}
{\tilde{D}}\,\epsilon ={\frac{{f^{2}}}{\rho }}\,D\,\omega +\psi \,{\tilde{D}}%
\,\chi .  \label{eps}
\end{equation}

The use of these potentials $\chi $ and $\epsilon $, will be helpful in the
procedure of defining harmonic functions, so we rewrite the field equations
in terms of these functions: 
\begin{eqnarray}
D^{2}\,\kappa +({\frac{{D\,\rho }}{\rho }}-{\frac{{D\,\kappa }}{\kappa }}%
)\,D\kappa +{\frac{{\alpha ^{2}\,\kappa ^{3}}}{{4\,f}}}(D\psi ^{2}-{\frac{1}{%
{\kappa ^{4}}}}\,D\,\chi ^{2}) &=&0,  \nonumber \\
D^{2}\,\psi +({\frac{{D\,\rho }}{\rho }}+{\frac{{2\,D\,\kappa }}{\kappa }}-{%
\frac{{D\,f}}{{f}}})\,D\,\psi -{\frac{1}{{\kappa ^{2}\,f}}}\,(D\,\epsilon
-\psi \,D\,\chi )\,D\,\chi &=&0,  \nonumber \\
D^{2}\,\chi +({\frac{{D\,\rho }}{\rho }}-{\frac{{2\,D\,\kappa }}{\kappa }}-{%
\frac{{D\,f}}{{f}}})\,D\,\chi +{\frac{{\kappa ^{2}}}{{f}}}\,(D\,\epsilon
-\psi \,D\,\chi )\,D\,\psi &=&0,  \nonumber \\
D^{2}\,f+({\frac{{D\,\rho }}{\rho }}-{\frac{{D\,f}}{{f}}})\,D\,f+{\frac{1}{{f%
}}}(D\,\epsilon -\psi \,D\,\chi )^{2}-{\frac{{\kappa ^{2}}}{{2}}}\,(D\psi
^{2}+{\frac{1}{{\kappa ^{4}}}}\,D\,\chi ^{2}) &=&0,  \nonumber \\
D^{2}\,\epsilon -D\,\psi \,D\,\chi -\psi \,D^{2}\chi +({\frac{{D\,\rho }}{%
\rho }}-{\frac{{2\,D\,f}}{{f}}})\,(D\,\epsilon -\psi \,D\,\chi ) &=&0,
\label{eqPot}
\end{eqnarray}
where we have used the fact that ${\tilde{D}}A\,{\tilde{D}}B=DA\,DB$, for
any functions $A,B$. The equation for $\chi $ is obtained from $D\,{\tilde{D}%
}\,A_{3}=0$, and the one for $\epsilon $ is obtained from $D\,{\tilde{D}}%
\,\omega =0$. Equations (\ref{eqPot}) are equivalent to equations (\ref
{diser}). This new set of field equations can be obtained from the
Lagrangian \cite{neu}, \cite{neu1} 
\begin{equation}
{\cal L}={\frac{\rho }{{2\,f^{2}}}}\,[D\,f^{2}+(D\,\epsilon -\psi \,D\,\chi
)^{2}]+{\frac{{2\,\rho }}{{\alpha ^{2}\,\kappa ^{2}}}}\,D\,\kappa ^{2}-{%
\frac{{\rho \,\kappa ^{2}}}{{2\,f}}}\,(D\psi ^{2}+{\frac{1}{{\kappa ^{4}}}}%
\,D\,\chi ^{2}).  \label{lag2}
\end{equation}
Notice, however, that this new Lagrangian is not obtained if the
transformation is made directly on the original Lagrangian, given by Eq.(\ref
{lag0}) or Eq.(\ref{lag1}). This fact implies that the transformations
defined by Eq.(\ref{chi}), and Eq.(\ref{eps}) must have a degeneracy. A
detailed analysis of this issue is under work and will be published
elsewhere.

As mentioned above, this Lagrangian, Eq.(\ref{lag2}), can be seen as given
by the line element, $DS^{2}$, of a potential space: ${\cal L}%
=DS^{2}=G_{AB}\,D\Psi ^{A}\,D\Psi ^{B}$, with $\Psi ^{A}=(f,\epsilon ,\chi
,\psi ,\kappa )$, so that the equations of motion, Eqs.(\ref{eqPot}),
obtained from variations of this Lagrangian with respect to the
``coordinates'', $\Psi ^{A}$, can be thought as ``geodesics in such
potential space''.

It is important to recall some of the geometric properties of this potential
space, $DS^{2}$. It defines a Riemannian space with constant scalar
curvature, $R=-(12+\alpha ^{2})$. All the covariant derivatives of the
Riemann tensor are proportional to $\alpha ^{2}-3$, thus the corresponding
Riemannian space is symmetric only for the case $\alpha ^{2}=3$ (and for the
case $\alpha =0$, not treated here). Therefore, for an arbitrary $\alpha $,
the space is not symmetric and thus the harmonic map ansatz formulation can
not be applied.

In this way, we see that in order to reformulate the
Einstein-Maxwell-Dilaton field equations for arbitrary $\alpha $, including
the low energy limit of string theory, we cannot use the harmonic map
formulation, but we can try instead to mimic that formulation, by means of
appropriately chosen ansatz. In this quest, it is necessary to obtain the
algebraic structure associated with the potential space. An analysis of the
killing vectors, $\xi ^{a}$, implies that there are 8 of them for the space
describe by Eq.(\ref{lag2}): 
\begin{eqnarray}
\xi _{1}^{a} &=&a_{1}\,(0,1,0,0,0),  \nonumber \\
\xi _{2}^{a} &=&a_{2}\,(0,0,1,0,0),  \nonumber \\
\xi _{3}^{a} &=&a_{3}\,(f,\epsilon ,{\frac{{\chi }}{{2}}},{\frac{{\psi }}{{2}%
}},0),  \nonumber \\
\xi _{4}^{a} &=&a_{4}\,(0,\chi ,0,1,0)  \nonumber \\
\xi _{5}^{a} &=&a_{5}\,(0,0,\chi ,-\psi ,\kappa )  \nonumber \\
\xi _{6}^{a} &=&a_{6}\,[(2\,\epsilon -\chi \,\psi )\,f,\epsilon
^{2}-f^{2}+f\,\psi ^{2}\,\kappa ^{2},  \nonumber \\
&&f\,\psi \,\kappa ^{2}+\epsilon \,\chi ,\psi \,\epsilon -\psi ^{2}\,\chi -{%
\frac{{f\,\chi }}{{\kappa ^{2}}}},{\frac{{\alpha ^{2}\,\psi \chi \kappa }}{{2%
}}}],  \nonumber \\
\xi _{7}^{a} &=&a_{7}\,[f\,\chi ,f\,\psi \,\kappa ^{2}+\epsilon \,\chi , 
\nonumber \\
&&f\,\kappa ^{2}+(1+\alpha ^{2})\,{\frac{{\chi ^{2}}}{{4}}},\epsilon
+(1-\alpha ^{2})\,{\frac{{\psi \,\chi }}{{2}}},\alpha ^{2}{\frac{{\,\chi
\,\kappa }}{{2}}}],  \nonumber \\
\xi _{8}^{a} &=&a_{8}\,[f\,\psi ,(3-\alpha ^{2}){\frac{{\chi \,\psi ^{2}}}{{4%
}}},(3-\alpha ^{2})\,{\frac{{\chi \,\psi }}{{2}}}-\epsilon ,  \nonumber \\
&&{\frac{{f}}{{\kappa ^{2}}}}+(1+\alpha ^{2})\,{\frac{{\psi ^{2}}}{{4}}}%
,-\alpha ^{2}\,{\frac{{\psi \,\kappa }}{{2}}}],
\end{eqnarray}
where $a_{1},...,a_{8}$, are arbitrary constants. For $\alpha ^{2}\neq 3$,
only the first 5 killing vectors remain independent. Thus, the algebra
associated with the potential space for arbitrary values of $\alpha $ is $%
SL(3,R)$ for $\alpha ^{2}=3,$ and a subalgebra of $SL(3,R)$, the upper
triangle one in the matrix representation, for $\alpha ^{2}\neq 3$.

The commutation relations, $\lbrack \xi^{a}_{i},\xi^{b}_{j}]= {\xi _{i}^{a}}%
_{;b}\,{\xi ^{b}}_{j}-{\xi _{j}^{a} }_{;b}\,{\xi ^{b}}_{j}$ for the first
five killing vectors are: 
\begin{eqnarray}
\lbrack \xi _{1},\xi _{2}] &=&0,\ \,\,\,\, [\xi _{1},\xi _{3}]=-\xi _{1},\
[\xi _{1},\xi _{4}]=0, \,\,\,\, [\xi _{1},\xi _{5}]=0,  \nonumber \\
\lbrack \xi _{2},\xi _{3}] &=&-{\frac{1}{{2}}}\xi _{2},\ [\xi _{2},\xi
_{4}]=-\xi _{1},\ [\xi _{2},\xi _{5}]=-\xi _{2},  \nonumber \\
\lbrack \xi _{3},\xi _{4}] &=&-{\frac{1}{{2}}}\xi _{4},\ [\xi _{3},\xi
_{5}]=0,\ \,\,\,\,[\xi _{4},\xi _{5}]=\xi _{4}.
\end{eqnarray}
Notice that we have several subalgebras with three vectors, but all of them
have at least one of the commutators equal to zero. These results will be
used in studying the type of target space that can be chosen in doing a map
to a bi-dimensional space.

Finally, as we mentioned, the remaining metric function $k$ is determined by
quadratures in terms of the other field functions \cite{MNQ}, explicitly: 
\begin{eqnarray}
k_{\rho } &=&{\frac{\rho }{{4\,f^{2}}}}[{f_{\rho }}^{2}-{f_{z}}^{2}+{%
\epsilon _{\rho }}^{2}-{\epsilon _{z}}^{2}+(\psi^2+{\frac{f}{\kappa ^{2}}}%
)\, ({\chi _{\rho }}^{2}-{\chi _{z}}^{2})-2\,\psi (\epsilon _{\rho }\,\chi
_{\rho }-\epsilon _{z}\,\chi _{z})+ \\
&&+f\,\kappa ^{2}({\psi _{\rho }}^{2}-{\psi _{z}}^{2})+({\frac{{2\,f}}{{%
\alpha \,\kappa }}})^{2}\,({\kappa _{\rho }}^{2}-{\kappa _{z}}^{2})], \\
k_{z} &=&{\frac{\rho }{{2\,f^{2}}}}[f_{\rho }\,f_{z}+\epsilon _{\rho
}\,\epsilon _{z}+\kappa ^{2}\,f\,\psi _{\rho }\,\psi _{z}-\psi \,(\epsilon
_{\rho }\,\chi _{z}+\epsilon _{z}\,\chi _{\rho })+ \\
&&(\psi^2+{\frac{f}{\kappa ^{2}}})\,\chi _{\rho }\,\chi _{z}+({\frac{{2\,f}}{%
{\alpha \,\kappa }}})^{2}\,\kappa _{\rho }\,\kappa _{z}].  \label{eq:k}
\end{eqnarray}

Continuing with the quest for reformulating the Lagrangian, Eq. (\ref{lag2}%
), notice that we do can perform a Legendre transformation and defining
``momenta'', as $P_{a}={\frac{{\partial {\cal L}}}{{\partial \,D\,\Psi ^{a}}}%
}$, we can construct a new function which is along the lines of the standard
Hamiltonian although in our case, it has not the usual properties of
evolution associated with the Hamiltonians. This ``Hamiltonian'' has the
explicit form: 
\begin{equation}
{\cal H}={\frac{{f^{2}}}{{2\,\rho }}}\,({P_{f}}^{2}+{P_{\epsilon }}^{2})+{%
\frac{{\alpha ^{2}\,\kappa ^{2}}}{{8\,\rho }}}\,{P_{\kappa }}^{2}-{\frac{{f}%
}{{2\,\rho }}}\,[{\frac{{{P_{\psi }}^{2}}}{{\kappa ^{2}}}}+\kappa ^{2}\,({%
P_{\chi }}+\psi \,{P_{\epsilon }})^{2}].  \label{ham}
\end{equation}

The equations of motion are now, $D\Psi ^{a}={\frac{{\partial {\cal H}}}{{%
\partial \,P_{a}}}}$, and $D\,P_{a}=-{\frac{{\partial {\cal H}}}{{\partial
\,\Psi ^{a}}}}$, that is: 
\begin{eqnarray*}
D\,f &=&{\frac{{f^{2}}}{{\rho }}}\,P_{f}, \\
D\,\epsilon &=&{\frac{{f^{2}}}{{\rho }}}\,[P_{\epsilon }-{\frac{{\kappa
^{2}\,\psi }}{{f}}}\,({P_{\chi }}+\psi \,{P_{\epsilon }})], \\
D\,\psi &=&-{\frac{{f}}{{\rho \,\kappa ^{2}}}}\,P_{\psi }, \\
D\,\chi &=&-{\frac{{f\,\kappa ^{2}}}{{\rho }}}\,({P_{\chi }}+\psi \,{%
P_{\epsilon }}), \\
D\,\kappa &=&{\frac{{\alpha ^{2}\,\kappa ^{2}}}{{4\,\rho }}}\,P_{\kappa },
\end{eqnarray*}
and 
\begin{eqnarray}
D\,P_{f} &=&-{\frac{{f}}{{\rho }}}\,({P_{f}}^{2}+{P_{\epsilon }}^{2})+{\frac{%
1}{{2\,\rho }}}\,[{\frac{{{P_{\psi }}^{2}}}{{\kappa ^{2}}}}+\kappa ^{2}\,({%
P_{\chi }}+\psi \,{P_{\epsilon }})^{2}],  \nonumber \\
D\,P_{\epsilon } &=&0,  \nonumber \\
D\,P_{\psi } &=&{\frac{{f\,\kappa ^{2}}}{{\rho }}}\,({P_{\chi }}+\psi \,{%
P_{\epsilon }})\,P_{\epsilon },  \nonumber \\
D\,P_{\chi } &=&0,  \nonumber \\
D\,P_{\kappa } &=&-{\frac{{f}}{{\rho \,\kappa }}}\,[{\frac{{{P_{\psi }}^{2}}%
}{{\kappa ^{2}}}}-\kappa ^{2}\,({P_{\chi }}+\psi \,{P_{\epsilon }})^{2}]-{%
\frac{{\alpha ^{2}\,\kappa }}{{4\,\rho }}}\,{P_{\kappa }}^{2}.  \label{eqH}
\end{eqnarray}
Now we can define new functions in order to simplify the form of this last
set of equations, and to be able to introduce the harmonic maps: 
\begin{eqnarray*}
A &=&{\frac{{f}}{{2\,\rho }}}\,(P_{f}-i\,P_{\epsilon }), \\
B &=&{\frac{\sqrt{f}}{{2\,\rho }}}\,[{\frac{{P_{\psi }}}{{\kappa }}}%
-i\,\kappa \,({P_{\chi }}+\psi \,{P_{\epsilon }})], \\
C &=&{\frac{{\alpha ^{2}\,\kappa }}{{4\,\rho }}}\,P_{\kappa },
\end{eqnarray*}
or in terms of the ``velocities'': 
\begin{eqnarray*}
A &=&{\frac{{1}}{{2\,f}}}\,[D\,f-i\,(D\,\epsilon -\psi \,D\,\chi )], \\
B &=&-{\frac{{1}}{{2\,\sqrt{f}}}}\,(\kappa \,D\,\psi -{\frac{i}{{\kappa }}}%
\,D\,\chi ), \\
C &=&{\frac{{D\,\kappa }}{{\kappa }}}\,
\end{eqnarray*}
so we can rewrite the equations for the ``momenta'', Eq.(\ref{eqH}) as the
following system: 
\begin{eqnarray}
{\frac{1}{{\rho }}}D(\rho \,A) &=&A\,(A-A^{\ast })+B\,B^{\ast },  \nonumber
\\
D(\rho \,B) &=&-{\frac{1}{{2}}}\,B(A-3\,A^{\ast })-C\,B^{\ast },  \nonumber
\\
D(\rho \,C) &=&-{\frac{{\alpha ^{2}}}{{2}}}\,(B^{2}+{B^{\ast }}^{2}),
\label{ABCeq}
\end{eqnarray}
where $^{\ast }$ denotes complex conjugate. Notice that in this way, we have
reduced the system of field equations to a set of three first order
differential equations for the three functions $A,B$, and $C$.

\section{The Ansatz}

{\hspace{1cm}}

Now we introduce the harmonic maps ansatz. Let $V_{p}$ be a $p$ dimensional
Riemannian space, and $\lambda ^{i}$ an harmonic parameters in $V_{p}$, $%
i.e. $%
\begin{equation}
(\rho \lambda _{,\zeta }^{i})_{,\bar{\zeta}}+(\rho \lambda _{,\bar{\zeta}%
}^{i})_{,\zeta }+2\rho \Gamma _{\ jk}^{i}\lambda _{,\zeta }^{j}\lambda _{,%
\bar{\zeta}}^{k}=0,
\end{equation}
where $\Gamma _{\ jk}^{i}$ are the Christoffel symbols on the Riemannian
space $V_{p}$. We suppose that all the potentials $f,\ \epsilon ,\dots $
depend on the $\lambda ^{i}$'s, $f=f(\lambda ^{i}),\ \epsilon =\epsilon
(\lambda ^{i}),\dots $ etc. (for details see ref. \cite{kra} \cite{ma24} and 
\cite{ma29}). We want to recall that this work belongs to a program aimed to
obtain exact solutions of the field equations derived from the Lagrangian (%
\ref{lag1}) by means of the harmonic maps ansatz (see \cite{ma24}) for the
case of arbitrary $\alpha $. However, as we mentioned, some difficulties
appear for arbitrary $\alpha $ and we cannot follow the procedures carried
out before. First, as was shown above, for $\alpha =0,\sqrt{3}$, the
isometry group has eight parameters, whereas for arbitrary $\alpha (\neq 0,%
\sqrt{3}),$ there exist only five parameters\ \cite{neu}. Second, the
isometry group of this Lagrangian is trivial except for $\alpha =0$ and $%
\alpha =\sqrt{3}$, in the sense that the invariant transformations of this
Lagrangian for arbitrary $\alpha $ lead only to gauge transformations.
Third, since for arbitrary $\alpha (\neq 0,\sqrt{3})$ the potential space is
not symmetric, it is not possible to obtain a no-linear sigma-model for this
system. All these problems prevent us to take the same way as in previous
works. Therefore we propose another method for solving the differential
equations. Here we will study the 2-dimensional subspace of the $V_{p}$%
-space. We start with a two dimensional Riemannian spaces $V_{2}$ with
constant curvature, parameterizing this Riemannian spaces with two harmonic
parameters $\lambda $, and $\tau $, such that $\lambda ,\tau \in %
\hbox{{I}\kern-.2em{\bf R}}$. The line element is 
\begin{equation}
ds^{2}={\frac{{2\,(d\,\lambda ^{2}\,+d\,\tau ^{2})}}{{(1-\sigma \,(\lambda
^{2}\,+\tau ^{2}))^{2}}}=}\frac{d\xi d\xi ^{\ast }}{(1-\sigma \xi \xi ^{\ast
})^{2}},  \label{le2d}
\end{equation}
where $\sigma $ is a real constant proportional to the potential space
curvature, and $\xi =\lambda +i\tau $, for the case of complex parameters.
We know that this is a maximally symmetric space, so it has three killing
vectors. If the electromagnetic field vanishes any value for $\alpha $ is
similar, because there is not interaction between scalar and electromagnetic
fields. Then $\sigma $ can be set to one as for $\alpha ^{2}=0,3$. But if
there is electromagnetic interaction the situation is different. As we
showed, the subalgebras for the potential space with arbitrary $\alpha $,
with three Killing vectors, are such that one of them has to be set to zero,
so we conclude that if the electromagnetic field does not vanish the only
case of maximally symmetric $V_{2}$ that can be taken is the one with $%
\sigma =0$. In this case, the parameters satisfy the usual Laplace equation: 
$D(\rho \,D\,\lambda )=0,$ $D(\rho \,D\,\tau )=0.$ As in the harmonic maps
ansatz case, let us express the functions $A,B,C$ in terms of these
parameters as follows: 
\begin{eqnarray}
A &=&a_{1}(\lambda ,\tau )\,D\,\lambda +a_{2}(\lambda ,\tau )\,D\,\tau , 
\nonumber \\
B &=&b_{1}(\lambda ,\tau )\,D\,\lambda +b_{2}(\lambda ,\tau )\,D\,\tau , 
\nonumber \\
C &=&c_{1}(\lambda ,\tau )\,D\,\lambda +c_{2}(\lambda ,\tau )\,D\,\tau .
\label{ABC}
\end{eqnarray}
Using the harmonic equations, $i.e.$ the Laplace equation, for these
parameters in the field equations (\ref{ABCeq}), and recalling the fact that 
$(D\,\lambda )^{2},(D\,\tau )^{2}$, and $D\,\lambda \,D\,\tau $ are
independent func\-ti\-ons, from the system of equations for $A,B,C$, Eq.(\ref
{ABCeq}), we obtain the following set of equations: 
\begin{eqnarray}
{a_{1}}_{,\lambda }-a_{1}\,(a_{1}-{a_{1}}^{\ast })-b_{1}\,{b_{1}}^{\ast }\,
&=&0,  \nonumber \\
{b_{2}}_{,\lambda }+{\frac{{b_{1}}}{2}}\,(a_{1}-3\,{a_{1}}^{\ast })+c_{1}\,{%
b_{1}}^{\ast } &=&0,  \nonumber \\
{c_{1}}_{,\lambda }+{\frac{{\alpha ^{2}}}{2}}\,({b_{1}}^{2}+{{b_{1}}^{\ast }}%
^{2})\, &=&0,  \label{abc1}
\end{eqnarray}
\begin{eqnarray}
{a_{2}}_{,\tau }-a_{2}\,(a_{2}-{a_{2}}^{\ast })-b_{2}\,{b_{2}}^{\ast } &=&0,
\nonumber \\
{b_{2}}_{,\tau }+{\frac{{b_{2}}}{2}}\,(a_{2}-3\,{a_{2}}^{\ast })+c_{2}\,{%
b_{2}}^{\ast }\, &=&0,  \nonumber \\
{c_{2}}_{,\tau }+{\frac{{\alpha ^{2}}}{2}}\,({b_{2}}^{2}+{{b_{2}}^{\ast }}%
^{2})\, &=&0,  \label{abc2}
\end{eqnarray}
\begin{eqnarray}
{a_{1}}_{,\tau }+{a_{2}}_{,\lambda }-2\,a_{1}\,a_{2}+{a_{1}}^{\ast
}\,a_{2}+a_{1}\,{a_{2}}^{\ast }-b_{1}\,{b_{2}}^{\ast }-{b_{1}}^{\ast
}\,b_{2} &=&0,  \nonumber \\
2\,{b_{1}}_{,\tau }+2\,{b_{2}}_{,\lambda }+b_{2}\,(a_{1}-3\,{a_{1}}^{\ast
})+b_{1}\,(a_{2}-3\,{a_{2}}^{\ast })+2\,c_{1}\,{b_{2}}^{\ast }+2\,c_{2}\,{%
b_{1}}^{\ast } &=&0,\newline
\nonumber \\
{c_{1}}_{,\tau }+{c_{2}}_{,\lambda }+\alpha ^{2}\,(b_{1}\,b_{2}+{b_{1}}%
^{\ast }\,{b_{2}}^{\ast }) &=&0.  \label{const}
\end{eqnarray}

Equations (\ref{abc1}, \ref{abc2}, \ref{const}) are equivalent to the field
equations (\ref{dieser2}). Taking the original potential also as functions
of the harmonic parameters, we can express the functions $A,B$, and $C$ from
Eq.(\ref{ABC}) as: 
\begin{eqnarray}
A &=&{\frac{{1}}{{2\,f}}}\,[f_{,\lambda }-i\,(\epsilon _{,\lambda }-\psi
\,\chi _{,\lambda })]\,D\,\lambda +{\frac{{1}}{{2\,f}}}\,[f_{,\tau
}-i\,(\epsilon _{,\tau }-\psi \,\chi _{,\tau })]\,D\,\tau ,  \nonumber \\
B &=&-{\frac{{1}}{{2\,\sqrt{f}}}}\,(\kappa \,\psi _{,\lambda }-{\frac{i}{{%
\kappa }}}\,\chi _{,\lambda })\,D\,\lambda -{\frac{{1}}{{2\,\sqrt{f}}}}%
\,(\kappa \,\psi _{,\tau }-{\frac{i}{{\kappa }}}\,\chi _{,\tau })\,D\,\tau ,
\nonumber \\
C &=&{\frac{{\kappa _{,\lambda }}}{{\kappa }}}\,D\,\lambda +{\frac{{\kappa
_{,\tau }}}{{\kappa }}}\,D\,\tau ,  \label{ABC2}
\end{eqnarray}
from which, and from Eq.(\ref{ABC}), we can make the following
identification: 
\begin{eqnarray}
a_{1}={\frac{{1}}{{2\,f}}}\,[f_{,\lambda }-i\,(\epsilon _{,\lambda }-\psi
\,\chi _{,\lambda })];\ \ &&a_{2}={\frac{{1}}{{2\,f}}}\,[f_{,\tau
}-i\,(\epsilon _{,\tau }-\psi \,\chi _{,\tau })],  \nonumber \\
b_{1}=-{\frac{{1}}{{2\,\sqrt{f}}}}\,(\kappa \,\psi _{,\lambda }-{\frac{i}{{%
\kappa }}}\,\chi _{,\lambda }); &&\,\ \ b_{2}=-{\frac{{1}}{{2\,\sqrt{f}}}}%
\,(\kappa \,\psi _{,\tau }-{\frac{i}{{\kappa }}}\,\chi _{,\tau }),  \nonumber
\\
c_{1}={\frac{{\kappa _{,\lambda }}}{{\kappa }}};\,\ \ \ \ \ &&c_{2}={\frac{{%
\kappa _{,\tau }}}{{\kappa }}}.  \label{iden}
\end{eqnarray}

\section{Classes of solutions}

Now we proceed to present some solutions to the Einstein-Maxwell-Dilaton
system in terms of the harmonic functions $\lambda $, $\tau $. Taking $%
a_{1},b_{1},c_{1}$ as functions of $\lambda $ only, and $a_{2},b_{2},c_{2}$
as functions of $\tau $ only, we obtain two sets of equations, one in terms
of $\lambda $, the other in terms of $\tau $, with the equations Eqs.(\ref
{const}), being only constraint equations. We will see three such cases:

\subsection{First Class}

Letting all the functions be complex, we obtain that $a_{2}=b_{2}=c_{2}=0$,
and 
\begin{eqnarray}
a_{1} &=&-\frac{1+i\sqrt{\gamma^{2}-\frac{1}{4}}}{2\,(\,\lambda +\beta _{0})}%
,  \nonumber \\
b_{1} &=&-\frac{\sqrt{\gamma}\,(\sqrt{\gamma -\frac{1}{2}} + i\,\sqrt{\gamma
+\frac{1}{2}})}{\alpha\,(\,\lambda +\beta _{0})},  \nonumber \\
c_{1} &=&-\frac{\gamma}{\lambda +\beta _{0}},  \label{sol11}
\end{eqnarray}
where $\beta _{0}$ is an arbitrary constant and $\gamma=\frac{\alpha}{2}\,%
\sqrt{\frac{3}{4-\alpha^{2}}}$. This solution is valid for $1< \alpha <2$.
Using Eqs.(\ref{iden}), we obtain that the potentials are given by: 
\begin{eqnarray}
f &=&\,\frac{f_{0}}{\lambda +\beta _{0}},  \nonumber \\
\epsilon &=&\frac{\epsilon _{0}}{\lambda +\beta _{0}}+ \epsilon
_{10}\,(\lambda +\beta_{0})^{-(\gamma+\frac{1}{2})} ,  \nonumber \\
\chi &=&\frac{2\,\kappa _{0}}{\alpha}\, \sqrt{\frac{f_0\,\gamma}{\gamma+%
\frac{1}{2}}}\, [(\lambda +\beta_{0})^{-(\gamma+\frac{1}{2})}-
\beta_{0}^{-(\gamma+\frac{1}{2})} ],  \nonumber \\
\psi &=&\frac{2}{\kappa _{0}\,\alpha}\, \sqrt{\frac{f_0\,\gamma}{\gamma-%
\frac{1}{2}}}\, [(\lambda +\beta_{0})^{\gamma-\frac{1}{2}}-
\beta_{0}^{\gamma-\frac{1}{2}}],  \nonumber \\
\kappa &=&\kappa _{0}\,(\,\lambda +\beta _{0})^{-\gamma },  \label{sol1}
\end{eqnarray}
where $\epsilon _{0}=f_{0}\, \sqrt{\frac{\gamma+\frac{1}{2}}{\gamma-\frac{1}{%
2}}}\, (\frac{4-\alpha^{2}}{\alpha^{2}}\,\gamma + \frac{1}{2})$, and . $%
\epsilon _{0}=-\frac{4\,f_0\,\gamma\,\beta_{0}^{\gamma-\frac{1}{2}}} {%
\alpha^2\,\sqrt{\gamma^2-\frac{1}{4}}}$ From this potentials, using Eqs.(\ref
{eps}, \ref{chi}), we obtain that 
\begin{eqnarray}
\omega _{,\rho } &=&-\frac{\sqrt{\gamma^{2}-\frac{1}{4}}}{f_{0}}\,\rho\,
\lambda_{,z}, \,\,\,\, \omega _{,z}=\frac{\sqrt{\gamma^{2}-\frac{1}{4}}}{%
f_{0}}\,\rho\, \lambda _{,\rho },  \nonumber \\
{A_{3}}_{,\rho } &=&-\frac{\sqrt{\gamma\,f_0}}{\kappa _{0}\,\alpha}\,
(\,\lambda +\beta _{0})^{\gamma -\frac{3}{2}}\, (\omega \,\sqrt{\gamma-\frac{%
1}{2}}\,\lambda _{,\rho}- \frac{\rho}{f_0}\,\sqrt{\gamma+\frac{1}{2}}\,
(\,\lambda +\beta _{0})\,\lambda _{,z}),  \nonumber \\
{A_{3}}_{,z} &=&-\frac{\sqrt{\gamma\,f_0}}{\kappa _{0}\,\alpha}\, (\,\lambda
+\beta _{0})^{\gamma -\frac{3}{2}}\, (\omega \,\sqrt{\gamma-\frac{1}{2}}%
\,\lambda _{,z}+ \frac{\rho}{f_0}\,\sqrt{\gamma+\frac{1}{2}}\, (\,\lambda
+\beta _{0})\,\lambda _{,\rho}).  \label{sol1AW}
\end{eqnarray}

Finally, from Eq.(\ref{eq:k}), we obtain that the last metric coefficient, $%
k $, in this case is constant, $k=k_{0}$, and we take $k_{0}=0$. This
solution represents a rotating object with scalar and electromagnetic fields
which depend on the form of $\lambda .$ In this way, we have obtained a
family of solutions for the Einstein-Maxwell-Dilaton theory in which all the
fields are non-trivially involved. Let us give an example. Writing the
metric (\ref{papa}) in spherical like coordinates $\rho =r\sin \theta ,$ $%
z=r\cos \theta $ , we can choose $\lambda =M/r$, which give us the line
element of the space-time as ($f_{0}=\beta _{0}=\frac{1}{2}$): 
\begin{equation}
ds^{2}=-\frac{1}{1+\frac{2\,M}{r}}(dt-2\,\sqrt{\gamma ^{2}-\frac{1}{4}}%
\,M\cos \theta d\phi )^{2}+(1+\frac{2\,M}{r})(dr^{2}+r^{2}d\theta
^{2}+r^{2}\sin \theta ^{2}d\phi ^{2}),  \label{eq:me}
\end{equation}
the scalar field is 
\begin{equation}
\varphi =\ln (\beta _{0}+\frac{M}{r})^{\frac{\gamma }{\alpha }}-\varphi _{0},
\end{equation}
where $\varphi _{0}=\ln \kappa _{0}^{\alpha }$, and the electric and
magnetic potentials are 
\begin{eqnarray}
A_{0} &=&\sqrt{\frac{\gamma }{\gamma -\frac{1}{2}}}\frac{[(1+\frac{2\,M}{r}%
)^{\gamma -\frac{1}{2}}-1]}{2^{\gamma }\,\kappa _{0}\alpha }, \\
A_{3} &=&\sqrt{\gamma (\gamma +\frac{1}{2})}\frac{M}{2^{\gamma -1}\,\kappa
_{0}\alpha }[(\beta _{0}+\frac{2\,M}{r})^{\gamma -\frac{1}{2}}-1]\cos \theta
.
\end{eqnarray}
Metric (\ref{eq:me}) contains a mass-like parameter, $M$, and is not
asymptotically flat, except for $\gamma ^{2}=\frac{1}{4}$. \ This space-time
represents a multipole electromagnetic field copled to a multiple
gravitational one. The electric charge is $q=\sqrt{\gamma \left( \gamma -%
\frac{1}{2}\right) }M/(2^{\gamma -1}\kappa _{0}\alpha )$, and magnetic
monopole charge $Q=\sqrt{\gamma \left( \gamma -\frac{1}{2}\right) }M\left(
\beta _{0}^{\gamma -\frac{1}{2}}-1\right) /\left( 2^{\gamma -1}\kappa
_{0}\alpha \right) $. Notice, however that the scalar field is tighted to
the gravitational and electromagnetic charges, in the sense that when $M=0,$
or $\gamma =0$, the scalar field becomes trivial. In order to overcome this
problem, we think that more potentials have to be included. Also, this
solution has no horizons, and is singular for $r=0$. Thus, we consider that
this solution is a good one for describing the space time with
gravitational, elemctromagnetic and dilatonic fields in a region between $%
r>0 $, and $r=r_{0}$ where it has to be matched to an exterior solution. A
detailed analysis of these physical properties of this particular exact
solution, as well as other solutions belonging to the family of solutions
presented in this section is underwork and will be published presently.

\subsection{Second Class}

We take the case where the ansatz is such that $%
a_{1},a_{2},b_{1},b_{2},c_{1},c_{2}\in \hbox{{I}\kern-.2em{\bf R}}$, and
that $a_{1},b_{1},c_{1}$depend only on $\lambda $, and $a_{2},b_{2},c_{2}$
depend only on $\tau $. With these conditions the system of equations, Eqs. (%
\ref{abc1}) implies that $b_{1}\,b_{2}=0$. Taking $b_{1}=0$, in turn implies
that $a_{1}=c_{1}={\rm const}$, and the final set of equations reduces to
the following: 
\[
{a_{2}}_{,\tau }-b_{2}^{2}=0,\newline
\,\,\,\,{b_{2}}_{,\tau }-{b_{2}}\,(a_{2}-c_{2})=0,\,\,\,\,\newline
{c_{2}}_{,\tau }+{\alpha ^{2}}\,{b_{2}}^{2}=0. 
\]
Solving for $a_{2}$, we get a Riccati type equation: 
\begin{equation}
{a_{2}}_{,\tau \tau }-(1+\alpha ^{2})({a_{2}}^{2})_{,\tau }+2\,k_{0}\,{a_{2}}%
_{,\tau }=0,
\end{equation}
with $k_{0}$ an integration constant. We obtain two sets of solutions for $%
a_{2}$: 
\begin{equation}
a_{2}={\frac{{k_{1}\,e^{q_{1}\,\tau }-k_{2}\,e^{(q_{1}+2(1+\alpha
^{2}))\,\tau }}}{{k_{1}\,e^{q_{1}\,\tau }+k_{2}\,e^{(q_{1}+2(1+\alpha
^{2}))\,\tau }}}}+{\frac{{k_{0}}}{{(1+\alpha ^{2})}}},  \label{sol2a2}
\end{equation}
and 
\begin{equation}
a_{2}={\frac{1}{{(1+\alpha ^{2})}}}\,(-{\frac{{k_{3}}}{{k_{3}\,\tau +k_{4}}}}%
+k_{0}),  \label{sol2a2,2}
\end{equation}
which imply 
\begin{eqnarray}
b_{2} &=&{\frac{{2\,\sqrt{-k_{1}\,k_{2}\,(1+\alpha ^{2})}\,e^{2(q_{1}+1+%
\alpha ^{2})\,\tau }}}{{k_{1}\,e^{q_{1}\,\tau }+k_{2}\,e^{(q_{1}+2(1+\alpha
^{2}))\,\tau }}}},  \nonumber \\
c_{2} &=&{\frac{{k_{0}}}{{(1+\alpha ^{2})}}}-\alpha ^{2}\,{\frac{{%
k_{1}\,e^{q_{1}\,\tau }-k_{2}\,e^{(q_{1}+2(1+\alpha ^{2}))\,\tau }}}{{%
k_{1}\,e^{q_{1}\,\tau }+k_{2}\,e^{(q_{1}+2(1+\alpha ^{2}))\,\tau }}}},
\label{sol2b2}
\end{eqnarray}
and 
\begin{eqnarray}
b_{2} &=&{\frac{{k_{3}}}{{(1+\alpha ^{2})^{\frac{1}{2}}\,(k_{3}\,\tau +k_{4})%
}}},  \nonumber \\
c_{2} &=&{\frac{{1}}{{(1+\alpha ^{2})}}}\,(k_{0}+{\frac{{\alpha ^{2}\,k_{3}}%
}{{k_{3}\,\tau +k_{4}}}}),  \label{sol2b2,2}
\end{eqnarray}
where $k_{i},q_{i}$ are constants. Performing the integration in equations (%
\ref{iden}), the following expressions for the potentials are obtained: 
\begin{eqnarray}
f &=&f_{0}\,{\frac{{e^{\lambda _{0}\,\lambda +\tau _{0}\,\tau }}}{{%
(m_{1}\,\Sigma _{1}+m_{2}\,\Sigma _{2})^{\gamma }}}},  \nonumber \\
\kappa ^{2} &=&{\kappa _{0}}^{2}\,(m_{1}\,\Sigma _{1}+m_{2}\,\Sigma
_{2})^{\beta }\,e^{\lambda _{0}\,\lambda +(\tau _{0}-t_{1}-t_{2})\,\tau }, 
\nonumber \\
\psi &=&{\frac{{m_{3}\,\Sigma _{1}+m_{4}\,\Sigma _{2}}}{{(m_{1}\,\Sigma
_{1}+m_{2}\,\Sigma _{2})}}},  \nonumber \\
\chi &=&0,  \nonumber \\
\epsilon &=&0,  \label{sol2pot}
\end{eqnarray}
and 
\begin{eqnarray}
f &=&{\frac{f_{0}\,e^{\gamma \,k_{1}\,\tau +\lambda _{0}\lambda }}{%
(k_{3}\tau +k_{4})^{\gamma }}},  \nonumber \\
\kappa ^{2} &=&{\kappa _{0}}^{2}\,{(k_{3}\tau +k_{4})^{\beta }}e^{\gamma
\,k_{1}\,\tau +\lambda _{0}\lambda }  \nonumber \\
\psi &=&-2f_{0}^{1/2}{\kappa _{0}\,(1+\alpha ^{2})^{1/2}\,(k_{3}\tau +k_{4})}%
,  \nonumber \\
\chi &=&0,  \nonumber \\
\epsilon &=&0,  \label{sol2pot2,2}
\end{eqnarray}
where $\gamma ={\frac{2}{{1+\alpha ^{2}}}}$, $\beta =\alpha ^{2}\,\gamma $, $%
f_{0},\kappa _{0},t_{i},\lambda _{0},\tau _{0},m_{i}$ are constants. For the
first set of solutions $\Sigma _{i}=e^{t_{i}\,\tau }$ and the constants are
related by 
\begin{equation}
4\,m_{1}\,m_{2}\,f_{0}+\kappa _{0}^{2}\,(1+\alpha
^{2})\,(m_{1}\,m_{4}-m_{2}m_{3})^{2}=0.  \label{sol2const}
\end{equation}
And for the second set of solutions $t_{1}=-t_{2}$, $\Sigma _{1}=\tau ,$ $%
\Sigma _{2}=1$, and the constants satisfy the relationship 
\begin{equation}
4\,m_{1}^{2}\,f_{0}-\kappa _{0}^{2}\,(1+\alpha
^{2})\,(m_{1}\,m_{4}-m_{2}m_{3})^{2}=0.  \label{sol2const2}
\end{equation}
In this way, with the ansatz chosen, we obtained families of solutions for
the Einstein-Maxwell-Dilaton theory, without magnetic field and without
rotation. The generic solutions are in terms of two arbitrary harmonic
functions and a large number of constants. The respectively magnetic
solutions can be obtained using the invariant trasformations $.$

\begin{equation}
\phi \rightarrow -\phi ,\,\,\,\, F_{\mu \nu }=1/2e^{-2\alpha \phi }\epsilon
_{\mu \nu \alpha \beta }F^{\alpha \beta }  \label{invtrans}
\end{equation}

Thus, we can generate particular solutions with very different physical
properties. These sets of solutions were presented as a rapid communication
in \cite{MNQ}, and it was shown that several interesting known solutions are
included, such as the static dilatonic version of the Kastor-Traschen
solution \cite{K-T}, and several generalizations of it; the spherically
symmetric dilatonic black hole, \cite{G-M}; and solutions with arbitrary
magnetic field coupled to the dilaton \cite{tona}.

In order to present explicitly some families of solutions, let us now
rewrite solutions (\ref{sol1}) in a more convenient form. If we perform the
following transformation, $f\rightarrow f/f_{0}$, $\kappa ^{2}\rightarrow
\kappa ^{2}/\kappa _{0}^{2}$, $\lambda \rightarrow \lambda _{0}\lambda +\tau
\tau _{0}$, $\tau _{0}\rightarrow -t_{1}-t_{2}$, and $g=m_{1}\,e^{t_{1}\,%
\tau }+m_{2}\,e^{t_{2}\,\tau }$ in (\ref{sol2pot}), then solution (\ref
{sol2pot}) transforms into 
\begin{eqnarray}
f &=&{\frac{e^{\lambda }}{g^{\gamma }}}\ ,\quad  \nonumber \\
\kappa ^{2} &=&{\frac{e^{-\lambda +\tau _{0}\tau }}{g^{\beta }}}\,  \nonumber
\\
\psi &=&\,{\frac{(m_{3}\Sigma _{1}+m_{4}\Sigma _{2})}{g}},\quad  \nonumber \\
\chi &=&0\,\quad  \nonumber \\
\epsilon &=&0,  \label{sol2}
\end{eqnarray}
\newline
or analogously for solution (\ref{sol2pot2,2}) we perform the transformation 
$f\rightarrow f/f_{0}$, $\kappa ^{2}\rightarrow \kappa ^{2}/\kappa _{0}^{2}$%
, $\lambda \rightarrow \lambda _{0}\lambda +\gamma \,k_{1}\tau $, $\psi
\rightarrow -\psi \,\,{\kappa _{0}(1+\alpha ^{2})^{1/2}}/f_{0}^{1/2}$, and $%
g=k_{3}\tau +k_{4}$ then solution (\ref{sol2pot2,2}) now reads 
\begin{eqnarray}
f={\frac{{e^{\lambda }}}{{\ g^{\gamma }}}},\quad &&\kappa ^{2}=e^{\lambda
}g^{\beta },  \nonumber \\
\psi ={\frac{1}{g}},\quad &&\chi =0,\,\,\,\,\epsilon =0.  \label{sol2,2}
\end{eqnarray}
In these two cases the differential equation for the metric function $k$ in (%
\ref{eq:k}) can be separated into a gravitational, a scalar and a
electromagnetic parts. In order to do so, we substitute (\ref{sol2}) and (%
\ref{sol2,2}) into (\ref{eq:k}) to obtain the differential equation for $k$,
thus arriving at 
\begin{equation}
k_{,\zeta }=\frac{\rho }{2}[{(\lambda _{,\zeta })^{2}+\frac{1}{\alpha ^{2}}%
((\lambda _{,\zeta }-\tau _{0}\tau _{,\zeta })^{2}-2q_{1}q_{2}\beta (\tau
_{,\zeta })^{2})}],  \label{ksepa}
\end{equation}
where $\zeta =\rho +i\,z$. Let us now perform the following separation 
\begin{equation}
k=k_{g}+k_{e}+k_{s},  \label{ksepara}
\end{equation}
where we have defined the gravitational part of $\ k$ as 
\begin{equation}
k_{g,\zeta }=\frac{\rho }{2}(\lambda _{,\zeta })^{2},\quad  \label{kg}
\end{equation}
the scalar part as 
\begin{equation}
k_{s,\zeta }=\frac{\rho }{2\alpha ^{2}}(\lambda _{,\zeta }-\tau _{0}\tau
_{,\zeta })^{2},  \label{ksc}
\end{equation}
and the electromagnetic part as 
\begin{equation}
k_{e,\zeta }=\frac{\rho }{\alpha ^{2}}q_{1}q_{2}\beta (\tau _{,\zeta })^{2},
\label{ke}
\end{equation}
where now the constants $a_{1},...,\ q_{1},q_{2},\ $and $\kappa _{0},$%
satisfy the relationship 
\begin{equation}
4a_{1}^{2}-\kappa _{0}^{2}(1+\alpha ^{2})(a_{1}a_{4}-a_{2}a_{3})^{2}=0.
\end{equation}

It is important to note that the electrostatic potential $\psi $ is
completely determined by the harmonic function $\tau $. This means that the
electrostatic (magnetostatic) potential is determined only by $\tau $, so we
can obtain solutions with arbitrary electromagnetic fields writing the
corresponding solution of the Laplace equation for $\tau $. The
corresponding magnetic solution can be obtained using the invariant
transformations (\ref{invtrans}). The most important well-know solutions can
be derived from this method. Some examples are given in \cite{rios}. New
solutions have been derived in \cite{ma32}, \cite{ma40} \cite{ma15}.

On the other hand, a star is basically a gravitational monopole together
with a magnetic dipole field. \ Using the invariant transformations (\ref
{invtrans}), we can construct a class of solution \ with this
characteristics. For the gravitational potential we now chose a
gravitational monopole, in order to reproduce the most important
gravitational features of the star and of the Schwarzschild solution. In
order to do so we write the line element (\ref{papa}) in Boyer-Lindquist
coordinates $\rho =\sqrt{r^{2}-2mr}\sin \theta ,$ $z=(r-m)\cos \theta $ and
take $\lambda =ln(1-\frac{2m}{r})$, the differential equation for $k_{g}$
can be integrated and the spacetime metric reads 
\begin{equation}
ds^{2}=e^{2(k_{s}+k_{e})}g^{\gamma }\frac{dr^{2}}{1-2m/r}+g^{\gamma
}r^{2}(e^{2(k_{s}+k_{e})}d\theta ^{2}+\sin ^{2}\theta d\varphi ^{2})-\frac{%
(1-2m/r)}{g^{\gamma }}dt^{2}  \label{met}
\end{equation}
the scalar and the electromagnetic fields for this solution read

\begin{eqnarray}
\kappa ^{2} &=&e^{-2\alpha \phi }=\frac{e^{-2\alpha \phi _{0}}}{%
(1-2m/r)g^{\beta }}  \nonumber \\
A_{3,\rho } &=&-Q\rho \tau _{,z}.,\,\,\,\,A_{3,z}=Q\rho \tau _{,\rho }
\label{metAK}
\end{eqnarray}
Metric (\ref{met}), (\ref{metAK}) is an static asymptotically flat metric
with mass parameter $m$, \ and magnetostatic charge $Q$. The scalar
parameter depends on the form of $\tau .$ Observe that the electromagnetic
field is always integrable because $\tau $ fulfills the Laplace equation $%
D(\rho D\tau )=(\rho \tau _{,z})_{,z}+(\rho \tau _{,\rho })_{,\rho }=0$. So
we can construct solutions representing magnetic monopoles, dipoles, etc.
With this form of the metric we can interpret $g$ and $e^{2k_{e}}$ as the
contribution\ of the electromagnetic field and $e^{2k_{s}}$ as the
contribution of the scalar field to the metric. Asymptotically, the scalar
field behaves like $e^{-2\alpha \phi _{0}}/g^{\beta }(1+2m/r+..)$, where $%
g\sim 1+O(1/r^{n}).$ This means that the scalar field deviate from a
constant in orders of \ $2m/r$. For a star like the sun $2m/r\sim 10^{-6},$
for a white dwarf $\ 2m/r\sim 10^{-4}$ and only for a compact star like a
neutron star $2m/r\sim 10^{-1},$ $i.e$, this metric represents the
space-time of an object very similar as a magnetized Schwarzschild one. Only
for a very compact object both metrics are different. This fact is in
agreement with the concept of spontaneous scalarization: if scalar fields
exist, compact stars will prefer to posses one in order to save energy, \
but even when a star posses one, it will very difficult to detect it (see 
\cite{ma}). We can carry out the follow classification of solution (\ref{met}%
), (\ref{metAK}):

\begin{enumerate}
\item  Suppose that $\tau =0$, this implies that $g=1$ and $A_{3}=0$, so
that the metric (\ref{met}) is 
\begin{equation}
ds^{2}=e^{2k_{s}}\frac{dr^{2}}{1-2m/r}+r^{2}(e^{2k_{s}}d\theta ^{2}+\sin
^{2}\theta d\varphi ^{2})-(1-2m/r)dt^{2}  \label{matos}
\end{equation}
This metric is almost spherically symmetric and represents a gravitational
body (gravitational monopole) with scalar field. The scalar field deforms
the spherical symmetry in the factor $e^{2k_{s}}d\theta ^{2}$. If $k_{s}=0$
we recover the spherical symmetry. This metric has been used as a model for
a star \cite{brena}, finding that the physical differences between (\ref
{matos}) and the Schwarzschild solution are too small to be measured, even
for a compact star like a pulsar. This implies that a star like the sun
could posses a scalar field and we will not be able to measure it.

\item  We now take $\tau \neq 0$, $g=1$ , but with arbitrary magnetic field $%
A_{3.}$ The metric (\ref{met}) then reads 
\begin{equation}
ds^{2}=e^{2(k_{s}+k_{e})}\frac{dr^{2}}{1-2m/r}+r^{2}(e^{2(k_{s}+k_{e})}d%
\theta ^{2}+\sin ^{2}\theta d\varphi ^{2})-(1-2m/r)dt^{2}
\end{equation}
In this case the scalar and the magnetostatic potentials are the ones that
deform the spherical symmetry of the metric. If we furthermore make $%
k_{s}+k_{e}=0$ and $\tau =\lambda $, $(p+q-1)^{2}-2pq\beta =0,$ we recover
the spherical symmetry and the metric transforms into the Schwarzschild line
element. This metric represents a gravitational body with arbitrary
magnetostatic field coupled to a scalar field. The scalar and mass
parameters here are proportional.

\item  If we choose $\tau =\lambda $, metric (\ref{met}) is spherically
symmetric, the constants fulfill the constrain $(p+q-1)^{2}-pq\beta =0$, and
the metric (\ref{met}) reads 
\begin{equation}
ds^{2}=g^{\gamma }\frac{dr^{2}}{1-2m/r}+g^{\gamma }r^{2}(d\theta ^{2}+\sin
^{2}\theta d\varphi ^{2})-\frac{(1-2m/r)}{g^{\gamma }}dt^{2}
\end{equation}
Here is contained the Gibbons- Maeda black hole \cite{G-M} by choosing $\tau 
$ to be the harmonic function corresponding to a monopole. The generalized
version of solution \cite{G-M} for any electromagnetic multipole field is
given by this metric.

\item  Finally, we choose $\tau =\lambda $ and $p+q=1$, which implies $%
k_{s}=0$. The metric (\ref{met}) is now given by 
\begin{equation}
ds^{2}=e^{2k_{e}}g^{\gamma }\frac{dr^{2}}{1-2m/r}+g^{\gamma
}r^{2}(e^{2k_{e}}d\theta ^{2}+\sin ^{2}\theta d\varphi ^{2})-\frac{(1-2m/r)}{%
g^{\gamma }}dt^{2}
\end{equation}
This metric is again almost spherically symmetric, but now it is deformed by
the electromagnetic field, the factor $e^{2k_{e}}d\theta ^{2}$ is the
deformation of the spherical symmetry due to the electromagnetic field.
\end{enumerate}

\subsection{Third Class}

Now let us study the case when the electromagnetic field vanishes. Here it
is convenient to take the harmonic parameter complex. Choosing the simple
ansatz $a_{2}=b_{1}=b_{2}=c_{1}=c_{2}=0,$ $\sigma =1$, using Eq.(\ref{le2d})
with the complex parameter $\xi$, and repeating the procedure, we obtain
that the system of equations Eqs. (\ref{abc1}), (\ref{abc2}) and (\ref{const}%
) takes the form 
\begin{eqnarray*}
{a_{1}}_{,\xi }-a_{1}^{2}-{\frac{{2\,\xi ^{\ast }\,a_{1}}}{{1-\xi \,\xi
^{\ast }}}} &=&0, \\
{a_{1}}_{,\xi ^{\ast }}+a_{1}\,a_{1}^{\ast } &=&0,
\end{eqnarray*}
with solution 
\begin{equation}
a_{1}={\frac{{1+\xi ^{\ast }}}{{(1+\xi )\,(\xi \,\xi ^{\ast }-1)}}}.
\end{equation}
Performing the integration to obtain the potential functions, one finally
gets 
\begin{eqnarray}
f &=&{\frac{{\xi \,\xi ^{\ast }-1}}{{(1+\xi )\,(1+\xi ^{\ast })}}}, 
\nonumber \\
\epsilon &=&i\,{\frac{{\xi -\xi ^{\ast }}}{{(1+\xi )\,(1+\xi ^{\ast })}}},
\end{eqnarray}
which we identify with the Ernst solution, that is, we have obtained the
To\-mi\-mat\-su-Sato family of solutions, including of course Kerr as a
particular case \cite{kra}.

\subsection{Fourth Class}

Finally, we obtain some stationary solutions using this method. The most
simple ansatz to solve equations (\ref{abc1}), (\ref{abc2}) and (\ref{const}%
) is supposing that all the functions $a_{1},a_{2},\dots $ etc. are
constants, the differential equations become then algebraic equations which
can be easily solved. For this case we take the ansatz 
\begin{eqnarray}
a_{1} &=&ip,\,\,\,a_{2}=0,  \nonumber \\
b_{1} &=&p(1-i),\,\,\,b_{2}=0,  \nonumber \\
d_{1} &=&-2p,\,\,\,c_{2}=0,
\end{eqnarray}
where $p$ is and arbitrary constant. Integrating the potentials for this
case we obtain 
\begin{eqnarray}
f &=&1,  \nonumber \\
\kappa ^{2} &=&{\kappa _{0}}^{2}\,e^{-4p\,\lambda },  \nonumber \\
\psi &=&-\frac{e^{2p\lambda }}{\kappa _{0}}+\psi _{0},  \nonumber \\
\chi &=&\kappa _{0}\,e^{-2p\,\lambda }+\chi _{0},  \nonumber \\
\epsilon &=&\kappa _{0}\,\psi _{0}\,e^{-2p\,\lambda }.
\end{eqnarray}
Substituting in the metric (\ref{papa}) and in the equation for the function 
$k$, (\ref{eq:k}), the metric can be integrated to obtain 
\begin{equation}
ds^{2}=-\,(dt-\omega d\,\varphi )^{2}+[e^{2k}(d\rho ^{2}+dz^{2})+\rho
^{2}d\varphi ^{2}]\ ,
\end{equation}
where $\omega $ and the function $k$ can be integrated from the differential
equations: 
\begin{eqnarray}
\omega _{,\rho } &=&-2p\ \rho \lambda _{,z}\,\,\,\,\omega _{,z}=2p\ \rho
\lambda _{,\rho }  \nonumber \\
k_{,\zeta } &=&\frac{p^{2}(3\alpha ^{2}+4)}{\alpha ^{2}}\rho (\lambda
_{,\zeta })^{2}  \label{deg}
\end{eqnarray}
The integrability conditions for $\omega $ and $k$ are guaranteed because $%
\lambda $ fulfills the Laplace equation. Metric (\ref{deg}) represents a
rotating degenerated object (the gravitational potential $f=1$), with
electromagnetic and scalar potentials, where the coupling constant $\alpha $
between scalar and electromagnetic fields remains arbitrary. The particular
multipole development depends on the harmonic function $\lambda $. As an
example, if $\lambda =\cos \theta /r$, the solution represents a pure
magnetic monopole with a multipole electrostatic field, without
gravitational one. In this case $\omega =-2p\sin ^{2}\theta /r$, and the
magnetic field is proportional to it. This represents thus a magnetic
dipole, with an exponentially decaying scalar field.

\section{Conclusions}

{\hspace{1cm}} We have presented a detailed description of the functional
space formulation, joined with the harmonic maps one, in such a way that
starting with the Einstein-Hilbert action for the stationary axisymmetric
space-time, we obtained an effective Lagrangian for the field variables, by
means of a Legendre transformation obtained a ``Hamiltonian'' and with a
canonical transformation reduce the system of dynamical equations to three
first order coupled differential equations for three unknowns, $A,B,C$, two
complex and one real. The harmonic maps ansatz enabled us to reduce the
system of five second order partial differential equations (\ref{eqPot}) for
the Einstein-Maxwell-Dilaton system with two Killing vectors to a system of
five (two complex and one real) first order ordinary differential equations (%
\ref{ABCeq}). Using the harmonic maps ansatz \cite{ma24}, we rewrote that
system of equations in such a way that, for different ansatz, it allows to
generate large classes of solutions to the Einstein-Maxwell theory non
minimally coupled to a dilatonic field. In this way, we consider that we
have described a robust formulation which allows us to generate large
classes of solutions, placing us on the right track, we think, to obtain
exact solutions for astrophysical important cases, like the one describing a
compact charged rotating object surrounded by a scalar field which could be
useful for solving the dark matter problem at a galactic level. Actually, in 
\cite{fco, DM}, we have worked out this idea and the scalar field does is a
good candidate for the dark matter in the galactic halo of spiral galaxies.
We have presented several particular cases which include most of the well
known solutions, as well as new ones, where we have explicitly presented the
form of the fields and of the charges, for particular choices of the
potentials. Also, the formulation described in the present work is not only
a technique to generate exact solutions, but also it has several other
possible directions worth studying, for instance, from the effective
Lagrangian (\ref{lag1}) an off-shell Lagrangian for the non-linear $\sigma $%
-model for some families of solutions can be obtained \cite{JDH}. It has not
been possible to make this analogy for cases with different values of $%
\alpha $, due to the fact that for them the potential space is not
symmetric. Finally we think that the ``Hamiltonian'' obtained in the present
formulation, is worth to study further within the gravity quantization
endeavor.

\section{Acknowledgments}

This work has been supported by DGAPA-UNAM, project IN121298, and by
CONACYT, Mexico, project 3697-E. TM thanks the hospitality from the
relativity group in Jena, Germany, and the DAAD support while this work was
partially done.

\end{document}